\documentclass[twocolumn,letterpaper,aps,prd,superscriptaddress,showpacs,floatfix,,nofootinbib,longbibliography]{revtex4-1}

\usepackage{graphicx}	% Include figure filed
\usepackage{amsmath}
\usepackage{longtable}
\usepackage{xspace}	% Include xspace

\newcommand{\pt}{\mbox{$p_T$}\xspace}

\newcommand{\absy}{\mbox{$|y|$}\xspace}
\newcommand{\meanpt}{\mbox{$\langle p_T \rangle$}\xspace}

\newcommand{\meanabsy}{\mbox{$\langle |y| \rangle$}\xspace}

\newcommand{\jpsi}{\mbox{$J/\psi$}\xspace}
\newcommand{\psip}{\mbox{$\psi^{\prime}$}\xspace}
\newcommand{\jpsiall}{\mbox{$A_{LL}^{J/\psi}$}\xspace}
\newcommand{\all}{\mbox{$A_{LL}$}\xspace}
\newcommand{\pp}{\mbox{$p$$+$$p$}\xspace}

\newcommand{\absetarange}{\mbox{$1.2<|\eta|<2.4$} for the north arm and \mbox{$1.2<|\eta|<2.2$} for the south arm\xspace}
\newcommand{\absyrange}{\mbox{$1.2<|y|<2.2$}\xspace}

\newcommand{\fbkg}{f_{\rm Bkg}}
\newcommand{\ahat}{$\hat a^{gg \rightarrow \jpsi + X}_{LL}$}

\newcommand{\specialcell}[2][c]{%to use when forcing line breaks
  \begin{tabular}[#1]{@{}c@{}}#2\end{tabular}}

\usepackage{multirow}
\usepackage{floatrow}
\begin{document}

\title{Measurements of double-helicity asymmetries in inclusive $J/\psi$ 
production in longitudinally polarized $p$$+$$p$ collisions 
at $\sqrt{s}=510$ GeV}

\newcommand{\abilene}{Abilene Christian University, Abilene, Texas 79699, USA}
\newcommand{\augie}{Department of Physics, Augustana University, Sioux Falls, South Dakota 57197, USA}
\newcommand{\banaras}{Department of Physics, Banaras Hindu University, Varanasi 221005, India}
\newcommand{\barc}{Bhabha Atomic Research Centre, Bombay 400 085, India}
\newcommand{\baruch}{Baruch College, City University of New York, New York, New York, 10010 USA}
\newcommand{\bnlcoll}{Collider-Accelerator Department, Brookhaven National Laboratory, Upton, New York 11973-5000, USA}
\newcommand{\bnlphys}{Physics Department, Brookhaven National Laboratory, Upton, New York 11973-5000, USA}
\newcommand{\caucr}{University of California-Riverside, Riverside, California 92521, USA}
\newcommand{\charlesczech}{Charles University, Ovocn\'{y} trh 5, Praha 1, 116 36, Prague, Czech Republic}
\newcommand{\chonbuk}{Chonbuk National University, Jeonju, 561-756, Korea}
\newcommand{\ciae}{Science and Technology on Nuclear Data Laboratory, China Institute of Atomic Energy, Beijing 102413, P.~R.~China}
\newcommand{\cns}{Center for Nuclear Study, Graduate School of Science, University of Tokyo, 7-3-1 Hongo, Bunkyo, Tokyo 113-0033, Japan}
\newcommand{\colorado}{University of Colorado, Boulder, Colorado 80309, USA}
\newcommand{\columbia}{Columbia University, New York, New York 10027 and Nevis Laboratories, Irvington, New York 10533, USA}
\newcommand{\czechtech}{Czech Technical University, Zikova 4, 166 36 Prague 6, Czech Republic}
\newcommand{\elte}{ELTE, E{\"o}tv{\"o}s Lor{\'a}nd University, H-1117 Budapest, P{\'a}zm{\'a}ny P.~s.~1/A, Hungary}
\newcommand{\ewha}{Ewha Womans University, Seoul 120-750, Korea}
\newcommand{\fsu}{Florida State University, Tallahassee, Florida 32306, USA}
\newcommand{\gsu}{Georgia State University, Atlanta, Georgia 30303, USA}
\newcommand{\hanyang}{Hanyang University, Seoul 133-792, Korea}
\newcommand{\hiroshima}{Hiroshima University, Kagamiyama, Higashi-Hiroshima 739-8526, Japan}
\newcommand{\howard}{Department of Physics and Astronomy, Howard University, Washington, DC 20059, USA}
\newcommand{\ihepprot}{IHEP Protvino, State Research Center of Russian Federation, Institute for High Energy Physics, Protvino, 142281, Russia}
\newcommand{\illuiuc}{University of Illinois at Urbana-Champaign, Urbana, Illinois 61801, USA}
\newcommand{\inrras}{Institute for Nuclear Research of the Russian Academy of Sciences, prospekt 60-letiya Oktyabrya 7a, Moscow 117312, Russia}
\newcommand{\instpasczech}{Institute of Physics, Academy of Sciences of the Czech Republic, Na Slovance 2, 182 21 Prague 8, Czech Republic}
\newcommand{\isu}{Iowa State University, Ames, Iowa 50011, USA}
\newcommand{\jaea}{Advanced Science Research Center, Japan Atomic Energy Agency, 2-4 Shirakata Shirane, Tokai-mura, Naka-gun, Ibaraki-ken 319-1195, Japan}
\newcommand{\jyvaskyla}{Helsinki Institute of Physics and University of Jyv{\"a}skyl{\"a}, P.O.Box 35, FI-40014 Jyv{\"a}skyl{\"a}, Finland}
\newcommand{\karoly}{K\'aroly R\'oberts University College, H-3200 Gy\"ngy\"os, M\'atrai \'ut 36, Hungary}
\newcommand{\kek}{KEK, High Energy Accelerator Research Organization, Tsukuba, Ibaraki 305-0801, Japan}
\newcommand{\korea}{Korea University, Seoul, 136-701, Korea}
\newcommand{\kurchatov}{National Research Center ``Kurchatov Institute", Moscow, 123098 Russia}
\newcommand{\kyoto}{Kyoto University, Kyoto 606-8502, Japan}
\newcommand{\lahorelums}{Physics Department, Lahore University of Management Sciences, Lahore 54792, Pakistan}
\newcommand{\lawllnl}{Lawrence Livermore National Laboratory, Livermore, California 94550, USA}
\newcommand{\losalamos}{Los Alamos National Laboratory, Los Alamos, New Mexico 87545, USA}
\newcommand{\lund}{Department of Physics, Lund University, Box 118, SE-221 00 Lund, Sweden}
\newcommand{\maryland}{University of Maryland, College Park, Maryland 20742, USA}
\newcommand{\mass}{Department of Physics, University of Massachusetts, Amherst, Massachusetts 01003-9337, USA}
\newcommand{\michigan}{Department of Physics, University of Michigan, Ann Arbor, Michigan 48109-1040, USA}
\newcommand{\muhlenberg}{Muhlenberg College, Allentown, Pennsylvania 18104-5586, USA}
\newcommand{\myongji}{Myongji University, Yongin, Kyonggido 449-728, Korea}
\newcommand{\nara}{Nara Women's University, Kita-uoya Nishi-machi Nara 630-8506, Japan}
\newcommand{\natmephi}{National Research Nuclear University, MEPhI, Moscow Engineering Physics Institute, Moscow, 115409, Russia}
\newcommand{\newmex}{University of New Mexico, Albuquerque, New Mexico 87131, USA}
\newcommand{\nmsu}{New Mexico State University, Las Cruces, New Mexico 88003, USA}
\newcommand{\ohio}{Department of Physics and Astronomy, Ohio University, Athens, Ohio 45701, USA}
\newcommand{\ornl}{Oak Ridge National Laboratory, Oak Ridge, Tennessee 37831, USA}
\newcommand{\orsay}{IPN-Orsay, Univ.~Paris-Sud, CNRS/IN2P3, Universit\'e Paris-Saclay, BP1, F-91406, Orsay, France}
\newcommand{\peking}{Peking University, Beijing 100871, P.~R.~China}
\newcommand{\pnpi}{PNPI, Petersburg Nuclear Physics Institute, Gatchina, Leningrad region, 188300, Russia}
\newcommand{\riken}{RIKEN Nishina Center for Accelerator-Based Science, Wako, Saitama 351-0198, Japan}
\newcommand{\rikjrbrc}{RIKEN BNL Research Center, Brookhaven National Laboratory, Upton, New York 11973-5000, USA}
\newcommand{\rikkyo}{Physics Department, Rikkyo University, 3-34-1 Nishi-Ikebukuro, Toshima, Tokyo 171-8501, Japan}
\newcommand{\saispbstu}{Saint Petersburg State Polytechnic University, St.~Petersburg, 195251 Russia}
\newcommand{\seoulnat}{Department of Physics and Astronomy, Seoul National University, Seoul 151-742, Korea}
\newcommand{\stonybrkc}{Chemistry Department, Stony Brook University, SUNY, Stony Brook, New York 11794-3400, USA}
\newcommand{\stonycrkp}{Department of Physics and Astronomy, Stony Brook University, SUNY, Stony Brook, New York 11794-3800, USA}
\newcommand{\tenn}{University of Tennessee, Knoxville, Tennessee 37996, USA}
\newcommand{\titech}{Department of Physics, Tokyo Institute of Technology, Oh-okayama, Meguro, Tokyo 152-8551, Japan}
\newcommand{\tsukuba}{Center for Integrated Research in Fundamental Science and Engineering, University of Tsukuba, Tsukuba, Ibaraki 305, Japan}
\newcommand{\vandy}{Vanderbilt University, Nashville, Tennessee 37235, USA}
\newcommand{\weizmann}{Weizmann Institute, Rehovot 76100, Israel}
\newcommand{\wigner}{Institute for Particle and Nuclear Physics, Wigner Research Centre for Physics, Hungarian Academy of Sciences (Wigner RCP, RMKI) H-1525 Budapest 114, POBox 49, Budapest, Hungary}
\newcommand{\yonsei}{Yonsei University, IPAP, Seoul 120-749, Korea}
\newcommand{\zagreb}{University of Zagreb, Faculty of Science, Department of Physics, Bijeni\v{c}ka 32, HR-10002 Zagreb, Croatia}
\affiliation{\abilene}
\affiliation{\augie}
\affiliation{\banaras}
\affiliation{\barc}
\affiliation{\baruch}
\affiliation{\bnlcoll}
\affiliation{\bnlphys}
\affiliation{\caucr}
\affiliation{\charlesczech}
\affiliation{\chonbuk}
\affiliation{\ciae}
\affiliation{\cns}
\affiliation{\colorado}
\affiliation{\columbia}
\affiliation{\czechtech}
\affiliation{\elte}
\affiliation{\ewha}
\affiliation{\fsu}
\affiliation{\gsu}
\affiliation{\hanyang}
\affiliation{\hiroshima}
\affiliation{\howard}
\affiliation{\ihepprot}
\affiliation{\illuiuc}
\affiliation{\inrras}
\affiliation{\instpasczech}
\affiliation{\isu}
\affiliation{\jaea}
\affiliation{\jyvaskyla}
\affiliation{\karoly}
\affiliation{\kek}
\affiliation{\korea}
\affiliation{\kurchatov}
\affiliation{\kyoto}
\affiliation{\lahorelums}
\affiliation{\lawllnl}
\affiliation{\losalamos}
\affiliation{\lund}
\affiliation{\maryland}
\affiliation{\mass}
\affiliation{\michigan}
\affiliation{\muhlenberg}
\affiliation{\myongji}
\affiliation{\nara}
\affiliation{\natmephi}
\affiliation{\newmex}
\affiliation{\nmsu}
\affiliation{\ohio}
\affiliation{\ornl}
\affiliation{\orsay}
\affiliation{\peking}
\affiliation{\pnpi}
\affiliation{\riken}
\affiliation{\rikjrbrc}
\affiliation{\rikkyo}
\affiliation{\saispbstu}
\affiliation{\seoulnat}
\affiliation{\stonybrkc}
\affiliation{\stonycrkp}
\affiliation{\tenn}
\affiliation{\titech}
\affiliation{\tsukuba}
\affiliation{\vandy}
\affiliation{\weizmann}
\affiliation{\wigner}
\affiliation{\yonsei}
\affiliation{\zagreb}
\author{A.~Adare} \affiliation{\colorado} 
\author{C.~Aidala} \affiliation{\michigan} 
\author{N.N.~Ajitanand} \affiliation{\stonybrkc} 
\author{Y.~Akiba}  \email[PHENIX Spokesperson: ]{akiba@bnl.gov} \affiliation{\riken} \affiliation{\rikjrbrc}
\author{R.~Akimoto} \affiliation{\cns} 
\author{M.~Alfred} \affiliation{\howard} 
\author{N.~Apadula} \affiliation{\isu} \affiliation{\stonycrkp} 
\author{Y.~Aramaki} \affiliation{\riken} 
\author{H.~Asano} \affiliation{\kyoto} \affiliation{\riken} 
\author{E.T.~Atomssa} \affiliation{\stonycrkp} 
% ...
\collaboration{PHENIX Collaboration} \noaffiliation

\date{\today}

%------------------------------------------------------------------------------|

\begin{abstract}
%\linenumbers

We report the double helicity asymmetry, $A_{LL}^{J/\psi}$, in inclusive 
$J/\psi$ production at forward rapidity as a function of transverse 
momentum $p_T$ and rapidity $|y|$. The data analyzed were taken during 
$\sqrt{s}=510$ GeV longitudinally polarized $p$$+$$p$ collisions at the 
Relativistic Heavy Ion Collider in the 2013 run using the PHENIX detector.  
At this collision energy, $J/\psi$ particles are predominantly produced 
through gluon-gluon scatterings, thus $A_{LL}^{J/\psi}$ is sensitive to 
the gluon polarization inside the proton.  We measured $A_{LL}^{J/\psi}$ 
by detecting the decay daughter muon pairs $\mu^+ \mu^-$ within the PHENIX 
muon spectrometers in the rapidity range $1.2<|y|<2.2$. In this kinematic 
range, we measured the $A_{LL}^{J/\psi}$ to be 
$0.012 \pm 0.010$~(stat)~$\pm$~$0.003$(syst). The $A_{LL}^{J/\psi}$ can be 
expressed to be proportional to the product of the gluon polarization 
distributions at two distinct ranges of Bjorken $x$: one at moderate range 
$x \approx 5\times 10^{-2}$ where recent data of jet and $\pi^0$ double 
helicity spin asymmetries have shown evidence for significant gluon 
polarization, and the other one covering the poorly known small-$x$ region 
$x \approx 2\times 10^{-3}$. Thus our new results could be used to further 
constrain the gluon polarization for $x< 5\times 10^{-2}$. \end{abstract}

	
\maketitle

%%%%%%%%%%%%%%%%%%%%%%%%%%%%%%%%%%%%%%
\section{Introduction}
%%%%%%%%%%%%%%%%%%%%%%%%%%%%%%%%%%%%%%

Understanding the proton spin structure in terms of quark and gluon 
degrees of freedom is one of the key open questions in the field of hadron 
physics. The total angular momentum of the proton may be decomposed into 
quark and gluon contributions in several different 
frameworks~\cite{Jaffe:1989jz,Ji:1996ek,Chen:2011gn,Hatta:2011zs,Ji:2013fga,Wakamatsu:2012ve}. 
For example, in the infinite momentum frame, the contributions to the 
proton spin can be classified according to the Manohar-Jaffe sum 
rule~\cite{Jaffe:1989jz,Ji:2014lra,Leader:2011za}:

\begin{equation}
\label{eq:Manohar_Jaffe_sum_rule}
S_p = \frac{1}{2} = \frac{1}{2}\Delta\Sigma + \Delta G + L_q + L_g .
\end{equation}

\noindent
Here, 1/2 $\Delta \Sigma$ represents the contribution from quark helicity 
distributions (quark polarization projected onto the proton momentum 
direction); similarly, $\Delta G$ represents the contribution from gluon 
helicity distributions; $L_q$ and $L_g$ represent the contributions from 
orbital angular momenta of quarks and gluons respectively. The 
Manohar-Jaffe scheme has been widely used to directly compare theoretical 
expectations with experimental data in the infinite momentum frame for 
quark and gluon polarization contributions; however, the direct connection 
between orbital angular momentum and any corresponding experimental 
observable is still under debate~\cite{Chen:2011gn, Wakamatsu:2012ve}.

The polarized parton distribution functions have been studied extensively 
at the European Laboratory for Particle Physics, the Standford 
Linear Accelerator, the Deutsches Elektronen-Synchrotron, 
the Thomas Jefferson National Accelerator Facility and 
the Relativistic Heavy Ion Collider (RHIC) for decades. The 
most-recent-global quantum-chromodynamics (QCD)  
fits~\cite{deFlorian:2008mr,deFlorian:2009vb,Leader:2010rb,Blumlein:2010rn,Hirai:2008aj,Nocera:2014gqa} 
based on these experimental data indicate that the quark polarization only 
accounts for about 30\% of the proton spin. The remaining spin must come 
from the contributions from gluon polarization and from the orbital 
angular momentum of quarks and gluons. To resolve this ``spin puzzle", it 
is critical to understand the contribution from gluon 
polarization~\cite{Altarelli:1988nr,Carlitz:1988ab,Efremov:1988zh,Anselmino:1988hn,Morii:1993ja}.

Many hard-scale processes in \pp collisions at RHIC energies are 
dominated by gluon-gluon and quark-gluon interactions; the corresponding 
spin observables are therefore sensitive to the gluon polarization. The 
latest global fits (DSSV~\cite{deFlorian:2014yva}, 
NNPDFpol~\cite{Nocera:2014gqa}, etc.) incorporating the RHIC 2009 
inclusive jet~\cite{Adamczyk:2014ozi} and $\pi^0$~\cite{Adare:2014hsq} 
spin asymmetry data at midrapidity show the first experimental evidence 
of sizable gluon polarization at moderate Bjorken $x$ in the range $0.05 
\leq x \leq 0.2$. With higher statistics, a recent PHENIX 
$A_{LL}^{\pi^{0}}$ measurement~\cite{Adare:2015ozj} extended the small $x$ 
reach down to $ 1\times10^{-2}$ for the polarized gluon distribution. 
However, in the smaller-$x$ region, $x < 1\times10^{-2}$, where gluons 
dominate, the gluon polarization remains poorly constrained.

The measurement of the double helicity asymmetry in the production of 
\jpsi particles at forward rapidity can provide access to the gluon 
polarization in a smaller $x$ region, $x \sim 2 \times 10^{-3}$. In $p+p$ 
collisions at RHIC energies, \jpsi particles are predominantly produced 
via gluon-gluon scatterings~\cite{Gupta:1997nj}. Therefore, at leading 
order, the asymmetry of \jpsi production can be expressed as:

\begin{align}
\label{eq:Jpsi_ALL_gluon}
\jpsiall &= \frac{\Delta \sigma}{\sigma} = \frac{\sigma^{++} - \sigma^{+-}}{\sigma^{++} + \sigma^{+-}}\\
&\approx \frac{\Delta g(x_{1})}{g(x_{1})}\otimes\frac{\Delta g(x_{2})}{g(x_{2})} \otimes \hat a^{gg \rightarrow \jpsi + X}_{LL}, 
\end{align}

\noindent
where \jpsiall is the \jpsi double helicity asymmetry defined by the ratio 
of the polarized and unpolarized \jpsi cross sections ($\Delta \sigma$ and 
$\sigma$); `$++$' and `$+-$' denote the same and opposite helicity \pp 
collisions; $\Delta g(x)$ and $g(x)$ are the polarized and unpolarized 
gluon parton distribution functions; and $\hat a^{gg \rightarrow \jpsi + 
X}_{LL}$ is the partonic double helicity asymmetry for the process of $g+g 
\rightarrow \jpsi + X$. Due to the large charm quark mass, perturbative 
QCD is expected to work for calculations of the \jpsi and other charmonia 
production cross sections in high energy deep inelastic scattering and 
$p+p$ collisions. The production mechanisms of charmonia have been studied 
extensively for decades, and several theoretical approaches, including 
nonrelativistic QCD (NRQCD), have been developed to describe various 
experimental observations~\cite{Lansberg:2006dh}. In high energy $p+p$ 
collisions, the individual partonic double helicity asymmetry 
$\hat a^{gg \rightarrow \jpsi + X}_{LL}$ has been calculated in 
perturbative QCD for both color-singlet and color-octet mechanisms in the 
NRQCD framework, and used to calculate the inclusive 
\jpsiall~\cite{Gupta:1997nj,Teryaev:1996sr,Klasen:2003zn,Yujie:2015pri}.

\begin{figure}
\centering
\includegraphics[width=1.0\linewidth]{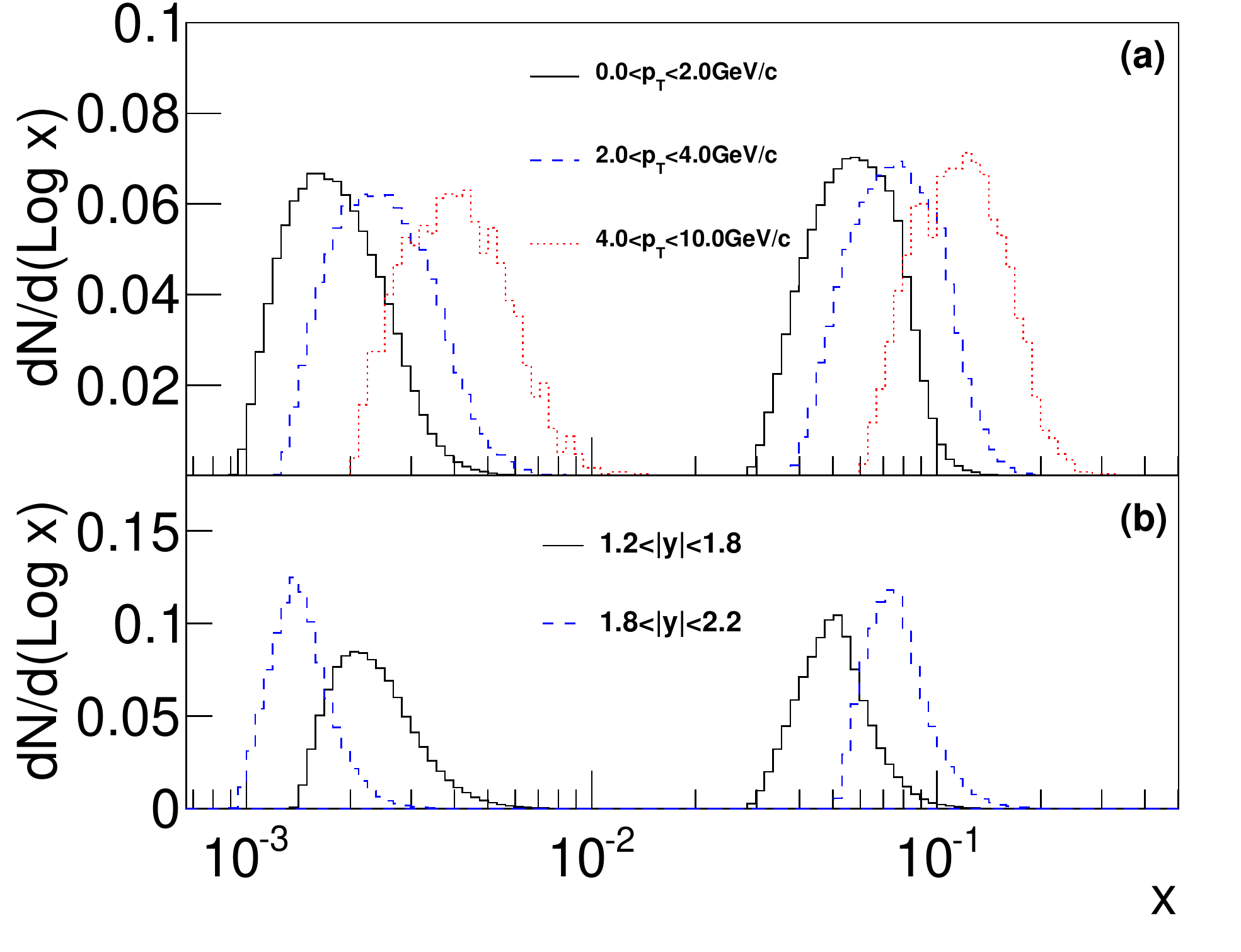}
\caption{\label{fig:x_dist} 
Bjorken $x$ distribution of gluons in the $gg \rightarrow J/\psi + X 
\rightarrow \mu^+ \mu^- + X$ process from a {\sc pythia} simulation with 
\jpsi generated within \absyrange and the decayed muon within 
\absetarange. Top panel shows the \pt binning and the bottom panel shows 
the \absy binning. All the distributions are arbitrarily normalized to 
have unit area.
}
\end{figure}

By detecting the \jpsi at forward rapidity, we sample participating gluons 
from two distinct ranges of Bjorken $x$. Quantitatively, we used a {\sc 
Pythia}~\cite{Sjostrand:PYTHIA} ({\sc pythia 6.4} tuned for RHIC energies) 
simulation at leading order to estimate the gluon $x$-distribution sampled 
in \jpsi production within the PHENIX muon arm acceptance. The simulation 
(Fig.~\ref{fig:x_dist}) illustrates that for the $g+g \rightarrow \jpsi + 
X$ process in the forward rapidity of the PHENIX muon arm acceptance, the 
two gluons come from two very distinct $x$ regions, with one gluon in the 
intermediate $x$ range ($3 \times 10^{-2}$ -- $2 \times 10^{-1}$) and the 
other gluon in the small $x$ range ($1 \times 10^{-3}$ -- $5 \times 10^{-3}$).

%%%%%%%%%%%%%%%%%%%%%%%%%%%%%%%%%%%%%%%%%%%%%%%%%%%%%%%%%%%%%%  Fig_1

Several sources contribute to the inclusive \jpsi production, including 
decays from heavier states containing charm and/or bottom quarks. Previous 
studies in PHENIX~\cite{Adare:2011vq} at midrapidity indicate that the 
excited states $\chi_{c}$ and \psip contribute a sizable (30\%--40\%) 
portion of the inclusive \jpsi production cross section. The $B 
\rightarrow \jpsi + X$ contribution is only important in the high $p_T > 
10$\,GeV region, and it is estimated to be small, less than 10\%
~\cite{Chatrchyan:2012np} in our kinematics at forward rapidity.

In the following, we present the measurement of the double helicity 
asymmetry in inclusive \jpsi production in longitudinally polarized \pp 
collisions at $\sqrt{s}$\,=\,510\,GeV. The data used for the study were 
collected by the PHENIX experiment~\cite{Adcox:2003zm} during the 2013 
run; the sampled integrated luminosity was about 150\,pb$^{-1}$ for this 
analysis.

%%%%%%%%%%%%%%%%%%%%%%%%%%%%%%%%%%%%%%
\section{Experiment Setup and Data Analysis}
%%%%%%%%%%%%%%%%%%%%%%%%%%%%%%%%%%%%%%

The \jpsi mesons were observed in the dimuon $\mu^+\mu^-$ decay channel 
using the two PHENIX forward muon spectrometers. Each spectrometer arm has 
full azimuthal coverage and spans the pseudorapidity range \absetarange. 
The major detector subsystems involved in this analysis were the muon 
trackers (MuTr) and the muon identifiers (MuID)~\cite{Akikawa:2003zs}, the 
beam-beam counters (BBC), the zero-degree calorimeters 
(ZDC)~\cite{Adler:2000bd}, and the forward-silicon-vertex detectors 
(FVTX)~\cite{Aidala:2013vna}.

The muon momentum was measured by the MuTr, a system based on three layers of 
cathode-strip tracking chambers in a radial-field magnet. The MuID 
comprises 5 layers of Iarocci tubes interleaved with 10 or 20 cm thick 
steel absorbers. The MuID absorbers, together with the central magnet 
absorbers (a combination of copper, iron and stainless steel, 
approximately 100 cm thick), were used to suppress light hadron 
backgrounds (pions and kaons) while allowing high energy muons to pass 
through. The probability of a high energy hadron ($p > 3$ GeV) generated 
from the interaction point (IP) passing through all the absorbers and 
getting mis-tagged as a muon is less than 3\%~\cite{Akikawa:2003zs} in 
$p+p$ collisions.

The BBC comprises two quartz \v{C}erenkov modules located on opposite sides 
of the IP at $z\,=\, \pm 144\,$cm, where z is the distance in the beam 
direction from the IP, and covering a pseudorapidity range of $3.1 < 
|\eta| < 3.9$ and full azimuth. The BBC system measures the collision 
vertex position along the beam direction via a time-of-flight method and 
also serves as one of the luminosity detectors.

Muon candidate events were selected using a BBC-based minimum-bias 
collision trigger in coincidence with a MuID track-based trigger. The MuID 
triggers were defined by various combinations of hits in several layers of 
the MuID projecting to the IP. A ``deep" MuID track requires at least one 
hit in the last two layers of the MuID detector and at least two hits in 
other layers. In the PHENIX 2013 run detector shielding configuration, a 
minimum momentum of $\sim$\,3\,GeV/$c$ was needed for muons to reach the 
last layer of the MuID. The data set we used was selected by the ``2-Deep 
Muon Trigger" which required at least two MuID deep tracks in the same 
muon arm in a $p+p$ collision event. A more detailed description of the 
2-Deep Muon Trigger is found in Ref.~\cite{Adare:2010bd}.

The ZDC detector comprises two hadron calorimeter arms at $|z| = 18\,$m. 
It covers a pseudorapidity range of $|\eta| > 6$. In this analysis, the 
ZDC served as a second luminosity detector for systematic studies.

The FVTX detector is composed of two end caps upstream of the 
MuTr~\cite{Aidala:2013vna}. By searching for common origin points of the 
detected tracks, the FVTX is capable of reconstructing primary collision 
vertices in the z range used in this measurement. The FVTX vertex 
resolution along the beam line direction is at the one millimeter level, 
which is much more precise than the vertex resolution of the BBC detector. 
In this analysis, the FVTX vertices were used when available to improve 
the mass resolution of the dimuon pairs.

%%%%%%%%%%%%%%%%%%%%%%%%%%%%%%%%%%%%%%%%%%%%%%%%%%%%%%%%%%%%%%  Fig_2
\begin{figure}[!htb]
\includegraphics[width=1.0\linewidth]{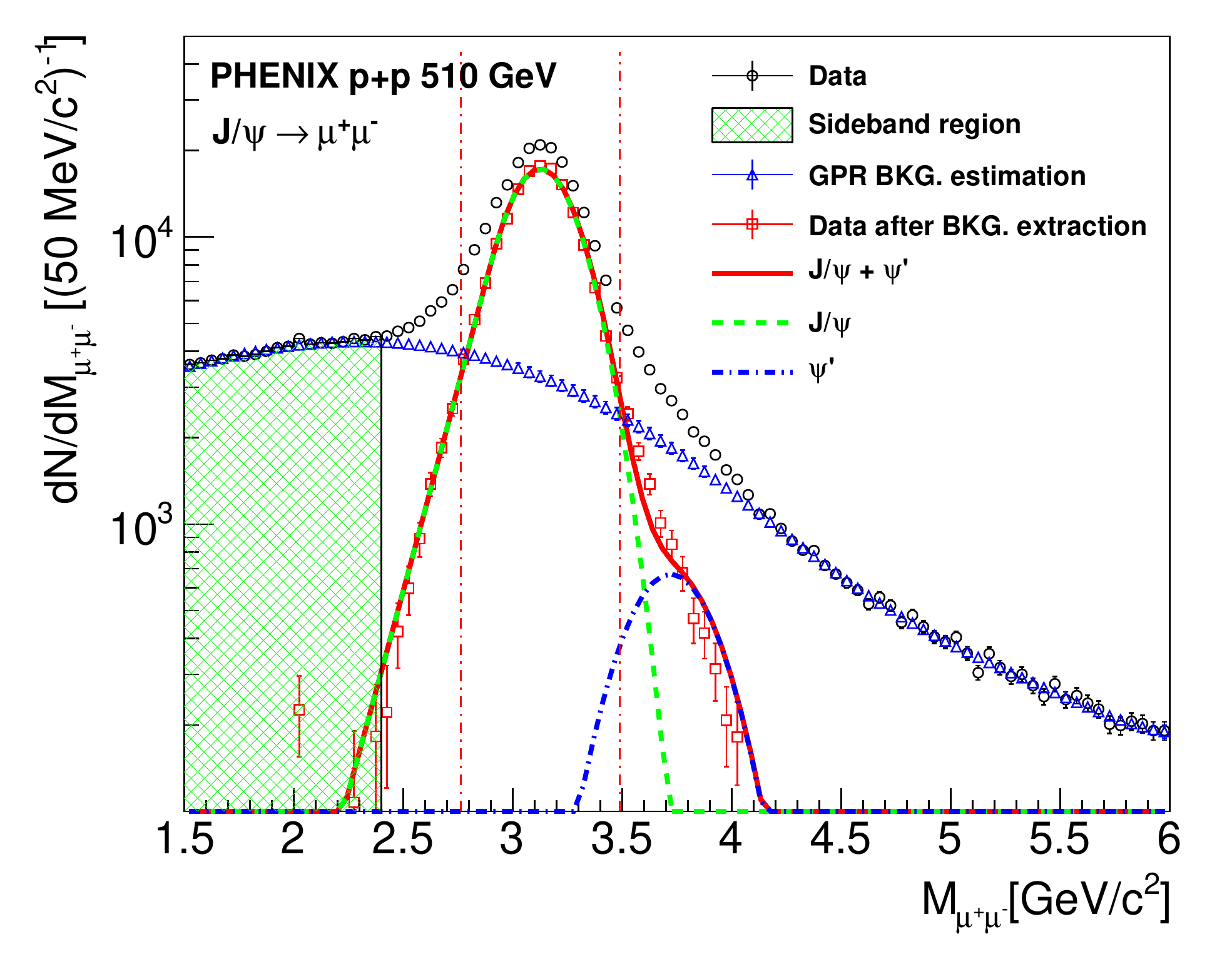}
\caption{\label{fig:fit} 
Dimuon invariant mass spectrum and the GPR fitting for the background 
fraction $\fbkg$ extraction. The black circles are the PHENIX unlike-sign 
dimuon data after event and track selection. The blue triangles are the 
GPR background estimation. The red boxes are the data remaining after 
subtraction of the background. The green dashed line represents the \jpsi 
shape; the blue dot-dashed line represents the \psip shape; and the red 
solid line the summation of \jpsi and \psip. Green shaded region indicates 
the sideband area used for the calculation of $A_{LL}^{\rm Bkg}$ in 
Eq.~\ref{eq:ALL_phys}. The data in the region between the two red vertical 
lines are the data used to calculate $A_{LL}^{\rm Incl}$ in 
Eq.~\ref{eq:ALL_phys}.}
\end{figure}

For optimal use of the muon spectrometers, the collision vertex 
reconstructed by the BBC was required to be within $\pm$\,30\,cm of the IP 
along the beam direction. Each muon track candidate was required to have a 
longitudinal momentum $p_z < 100$ GeV/$c$ and transverse momentum 
\pt\,$<$\,10\,GeV/$c$. The distance between the projected MuTr track and 
MuID track position at the first layer of the MuID plane was required to 
be less than 15 cm, and the projected opening angle between the MuTr track 
and the MuID track less than 10 degrees. Similar MuTr and MuID track 
matching cuts were used in Ref.~\cite{Adare:2010bd}. A fit to the common 
vertex of the two candidate tracks near the IP was performed and was 
required to have a $\chi^2 < 20$ for 4 degrees of freedom. The black 
circle data points in Fig.~\ref{fig:fit} show the invariant mass 
distribution of the unlike-sign dimuon pairs after event and track quality 
selections.

At RHIC, the clockwise (``Blue") and counter-clockwise (``Yellow") 
circulating beams collide at several fixed IPs, the PHENIX detector being 
one of them. During the 2013 longitudinally polarized $p+p$ run, up to 111 
radio-frequency bunches in each beam were filled with protons. Protons in 
each bunch were configured to have positive or negative helicity, denoted 
as ``$+$" or ``$-$". Thus collisions at the PHENIX IP can be categorized 
into two helicity combinations: same helicity (denoted as $++$) and 
opposite helicity (denoted as $+-$) collisions. For parity-conserving QCD 
processes, the production cross sections obey the relations $\sigma^{++} = 
\sigma^{--}$ and $\sigma^{+-}=\sigma^{-+}$. Experimentally, the double 
helicity asymmetry is defined as:

\begin{equation}
\label{eq:A_LL_formula}
\begin{split}
A_{LL} &= \frac{\sigma^{++} - \sigma^{+-}}{\sigma^{++}+\sigma^{+-}} \\
 &= \frac{1}{P_{B}P_{Y}} \frac{N^{++}-R \cdot N^{+-}}{N^{++}+R \cdot N^{+-}},
\end{split}
\end{equation}

\noindent 
where $P_B$ ($P_Y$) is the beam polarization for the Blue (Yellow) beam, 
$\sigma^{++}$ ($\sigma^{+-}$) is the cross section for same (opposite) 
helicity collisions, and $N^{++}$ ($N^{+-}$) is the produced dimuon yield 
for same (opposite) helicity collisions. $R$ is the relative luminosity 
between helicity states and is defined as

\begin{equation}
\label{eq:relative_L_R}
R = \frac{L^{++}}{L^{+-}},
\end{equation}

\noindent
where $L^{++} (L^{+-})$ is the luminosity measured by the BBC detectors in 
$++ (+-)$ helicity state collisions. The averaged polarizations for the 
data used in this analysis were:

\begin{align}
\label{eq:avg_pol}
& P_B = 0.55 \pm 0.02 \,({\rm syst}), \\
& P_Y = 0.56 \pm 0.02 \,({\rm syst}).
\end{align}

For each ``fill" (a unit of the operation period of the accelerator, 
typically several hours) of the rings, the helicity pattern was designed 
to provide almost equal numbers of collisions in the $++$, $+-$, $-+$, and 
$--$ helicity configurations. In this way, slow changes in detector 
acceptance and efficiency were eliminated from the asymmetry determination 
in Eq.~\ref{eq:A_LL_formula}.

As shown in Fig.~\ref{fig:fit}, there is a small amount ($\sim 15\%$) of 
dimuon background underneath the $J/\psi$ signal peak in the dimuon 
invariant mass distribution; the background events may have a different 
asymmetry from that of $J/\psi$ events. To correct for this, we estimated 
the background asymmetry using the ``sideband" in the invariant mass 
region (1.5\,--\,2.4\,GeV/$c^{2}$), the green shaded region in 
Fig~\ref{fig:fit}.  Consistent with Ref.~\cite{Adare:2010bd}, this 
sideband was located below the \jpsi peak in invariant mass; a sideband 
that was higher in invariant mass would need to be placed further away 
from the \jpsi to avoid the \psip and would have had negligible 
statistical significance. For the final \jpsi double helicity asymmetry, 
we subtracted the background contributions:

\begin{eqnarray}
\label{eq:ALL_phys}
A_{LL}^{J/\psi} = \frac{A_{LL} ^{\rm Incl} - \fbkg \cdot A_{LL}^{\rm Bkg}}{1 - \fbkg},
\end{eqnarray}

\noindent
where \all values on the right-hand-side were calculated using 
Eq.~\ref{eq:A_LL_formula}. The asymmetry $A_{LL}^{\rm Incl}$ is for 
inclusive unlike-charge dimuon pairs in the invariant mass region $\pm 
2\sigma$ around the \jpsi mass peak mean value ($\sigma$ is the mass 
resolution of the detector), and $A_{LL}^{\rm Bkg}$ is the asymmetry for a 
sideband of unlike-charge dimuon pairs. In this analysis, the measured 
$A_{LL}^{\rm Bkg}$ was $-0.002 \pm 0.012 (stat)$ for the \pt range $0 
<$~\pt~$< 10$~GeV. The background fraction $\fbkg$ is defined as:

\begin{eqnarray}
\label{eq:r_def}
\fbkg = \frac{N_{\rm Bkg}}{N_{\rm Incl}},
\end{eqnarray}

\noindent
where $N_{\rm Bkg}$ is the number of estimated non\jpsi dimuon pairs in 
the $\pm 2 \sigma$ range around the \jpsi peak, and $N_{\rm Incl}$ is the 
total number of unlike-charge dimuon pairs in the same mass range. For the 
background under the \jpsi mass peak, a Gaussian Process Regression 
(GPR)~\cite{MacKay,Rasmussen,Lauritzen, barber2012bayesian,Adare:2015gsd} 
approach was used to determine the background distribution. Two training 
zones, on either side of the \jpsi peak, were defined for this GPR 
approach: 1.5\,--\,2.2\,GeV/$c^{2}$ and 4.3\,--\,6.0\,GeV/$c^{2}$. These 
two training zones were used only for the estimation of background yield, 
not the background asymmetry. The \jpsi $2\sigma$ mass window was defined 
by fitting the data after the GPR background subtraction. In the fitting, 
the \jpsi invariant mass peak shape was described by a Crystal Ball 
distribution~\cite{Skwarnicki:1986xj}, and for simplicity the low 
statistics $\psip$ peak was fit with a Gaussian distribution with mass 
resolution evaluated from Monte Carlo simulation.

In this analysis, we measured the asymmetry separately for the two muon 
arms. The results were then cross-checked for consistency and combined to 
produce the final physics double helicity asymmetry.

To further study the \pt- or \absy-dependence of the asymmetry, the data 
were divided into three \pt bins (0\,--\,2\,GeV/$c$, 2\,--\,4\,GeV/$c$, 
4\,--\,10\,GeV/$c$) or two \absy bins (1.2\,--\,1.8, 1.8\,--\,2.2). 
\jpsiall was extracted for each of the bins following the procedure 
described above; the corresponding background fraction $\fbkg$ was 
extracted and is listed in Table~\ref{tab:bgfrac}.

The statistical uncertainties for \jpsiall ($\Delta A_{LL}^{J/\psi}$) were 
calculated via Eq.~\ref{eq:dALL_phys}:

\begin{eqnarray}
\label{eq:dALL_phys}
\Delta A_{LL}^{J/\psi} = \frac{\sqrt{(\Delta A_{LL} ^{\rm Incl})^2 + (\fbkg \cdot \Delta A_{LL}^{\rm Bkg})^2}}{1 - \fbkg},
\end{eqnarray}

\noindent
where $\Delta A_{LL} ^{\rm Incl}$ and $\Delta A_{LL}^{\rm Bkg}$ represent 
the statistical uncertainty of the $A_{LL} ^{\rm Incl}$ and $A_{LL}^{\rm 
Bkg}$ respectively. The statistical uncertainty of $\fbkg$ is combined 
with its systematic uncertainty from the extraction method and considered 
as one of the systematic uncertainties which is discussed in the next 
section.

%%%%%%%%%%%%%%%%%%%%%%%%%%%%%%%%%%%%%%%%%%%%%%%%%%%%%%%%%%%%%%  Table_I
\begin{table}[!htb]
\caption[$background fraction$s]
{Background fraction $\fbkg$ for each arm and each \pt or \absy bin using 
the corresponding \jpsi $2 \sigma$ mass window for that bin. The 
systematic uncertainty is 0.05 (absolute value) for all the bins; see 
discussion in the text.}
\label{tab:bgfrac}
\begin{ruledtabular}
\begin{tabular}{cc}
\pt or \absy range	&	$\fbkg\pm\Delta\fbkg\,(stat)$ \\ \hline
$0<$~\pt~$<2$~GeV/$c$			&	0.26 $\pm$ 0.01	\\
$2<$~\pt~$<4$~GeV/$c$			&	0.17 $\pm$ 0.01	\\
$4<$~\pt~$<10$~GeV/$c$			&	0.18 $\pm$ 0.01	\\
$1.2<$~\absy~$<1.8$				&	0.25 $\pm$ 0.02 	\\
$1.8<$~\absy~$<2.2$				&	0.30 $\pm$ 0.02	\\
\end{tabular}
\end{ruledtabular}
\end{table}

%%%%%%%%%%%%%%%%%%%%%%%%%%%%%%%%%%%%%%
\section{Systematic uncertainty}
%%%%%%%%%%%%%%%%%%%%%%%%%%%%%%%%%%%%%%

There are two types of systematic uncertainties involved in this analysis: 
Type A are uncorrelated point-to-point uncertainties for each \pt or \absy 
bin, and Type B are correlated point-to-point uncertainties.

One important Type A systematic uncertainty comes from the determination 
of the background fraction under the \jpsi mass peak. To test the possible 
bias of the background fraction $\fbkg$ extracted from the GPR procedure, 
we compared to the method that was used in ~\cite{Adare:2010bd} which used 
a third order polynomial to describe the background. The two methods 
differed at most by 0.05 (absolute value); we took that as the systematic 
uncertainty for the background fraction $\fbkg$.

Another Type A systematic uncertainty is from the determination of 
background asymmetry under the \jpsi mass peak. Because the low mass side 
band was used to estimate the background spin asymmetry under the \jpsi 
mass peak, we need to estimate the bias introduced by this approximation. 
We studied the mass dependence of the background asymmetry by dividing the 
side band into two mass bins, 1.5\,--\,2.0\,GeV/$c^{2}$ and 
2.0\,--\,2.4\,GeV/$c^{2}$. We found no obvious mass-dependence beyond 
expected statistical fluctuation. Thus we concluded that this systematic 
uncertainty related to the mass-dependence of the background asymmetry is 
small compared with the statistical uncertainty of the sideband dimuon 
asymmetry ($\Delta A_{LL}^{\rm Bkg}$ in Eq.~\ref{eq:dALL_phys}) and is not 
counted as additional uncertainty for this analysis.

The last Type A systematic uncertainty comes from the variation of 
detector efficiency within a data group in which the asymmetry is 
calculated. For the purpose of getting sufficient statistics in the 
asymmetry calculations using Eq.~\ref{eq:A_LL_formula} discussed above, we 
collected individual PHENIX DAQ runs into larger groups. Each DAQ run 
corresponds to a time period of up to 1.5~hour of continuous data 
acquisition. However, the detector efficiency may vary between runs in 
each group, and that could lead to a biased result. The muon 
reconstruction efficiency has a dependence on the luminosity and event 
vertex distribution and it could also change over time. To study this 
systematically, three grouping methods were applied and compared with each 
other: (1) runs with similar luminosity and event vertex distribution; (2) 
runs within a RHIC fill to minimize the time spreading of each group; (3) 
all the runs into one group. We chose method (1) results to calculate the 
mean value of our results. The systematic uncertainty from the grouping 
method was set to the maximum variation extracted from these three 
approaches. Type A systematic uncertainties for all \pt or \absy bins are 
summarized in Table~\ref{tab:type_a_syst}.

%%%%%%%%%%%%%%%%%%%%%%%%%%%%%%%%%%%%%%%%%%%%%%%%%%%%%%%%%%%%%%  Table_II
\begin{table}
\caption{\label{tab:type_a_syst}
Type A systematic uncertainties for each \pt or \absy bin. $\Delta 
A_{LL}^{\rm fit}$ is the systematic uncertainty from background fraction 
determination. $\Delta A_{LL}^{\rm run\,group}$ is the systematic 
uncertainty from the run grouping method.}
\begin{ruledtabular}
\begin{tabular}{ccc}
\pt or \absy range &  $\Delta A_{LL}^{\rm fit}$	&  $\Delta A_{LL}^{\rm run\,group}$ \\\hline
$0<$~\pt~$<2$~GeV/$c$   &$<$\,0.001  &0.003  \\
$2<$~\pt~$<4$~GeV/$c$   &0.001  &0.004  \\
$4<$~\pt~$<10$~GeV/$c$  &0.003  &0.009  \\
$1.2<$~\absy~$<1.8$   &0.005  &0.004  \\
$1.8<$~\absy~$<2.2$   &0.002  &0.002  \\
\end{tabular}
\end{ruledtabular}
\end{table}

The systematic uncertainty in the determination of the relative luminosity 
is of Type B. The luminosities $L^{++,+-}$, and therefore also the 
relative luminosity $R$ used in Eq.~\ref{eq:A_LL_formula}, were measured 
by the BBC trigger counts with a vertex cut of $\pm$\,30\,cm along the 
beam line. To test if the BBC count rate contains an unmeasured physics 
asymmetry, we used another luminosity detector, the ZDC, and computed the 
double helicity asymmetry of the ZDC/BBC luminosity ratio:

\begin{equation}
\label{eq:rellumi_zdc_bbc}
A_{LL} ^{ZDC/BBC} = \frac{1}{P_{B}P_{Y}} \frac
{\frac{N_{ZDC}^{++}}{N_{\rm BBC}^{++}} - \frac{N_{ZDC}^{+-}}{N_{\rm BBC}^{+-}}}
{\frac{N_{ZDC}^{++}}{N_{\rm BBC}^{++}} + \frac{N_{ZDC}^{+-}}{N_{\rm BBC}^{+-}}},
\end{equation}

\noindent
where $N_{ZDC}$ ($N_{\rm BBC}$) is the coincidence counts measured by the ZDC 
(BBC), which is proportional to the beam luminosity. During the 2013 
PHENIX 510\,GeV \pp run, due to high beam intensity, approximately 30\% of 
bunch crossings contain more than one \pp binary collision. However, 
neither the BBC nor the ZDC can separate these multiple collisions. 
Therefore, multiple collisions are counted as one \pp collision and this 
affects the determination of the relative luminosity. A statistical 
pile-up correction was performed to remove the bias of the (relative) 
luminosity measurement caused by multiple collisions, identical to the 
correction performed in Ref.~\cite{Adare:2015ozj}. We took the asymmetry 
$A_{LL} ^{ZDC/BBC}$ plus its statistical uncertainty as a systematic 
uncertainty for the relative luminosity $R$. After pile-up corrections the 
systematic uncertainty from relative luminosity was determined to be $4 
\times 10^{-4}$.

Another source of systematic uncertainty (Type B) comes from the 
measurement of the average beam polarizations, $P_B$ and $P_Y$. The 
uncertainty of the product $P_B P_Y$ used in Eq.~\ref{eq:A_LL_formula} 
leads to an overall scale uncertainty of the \all measurements. For the 
RHIC 2013 data set, this uncertainty was evaluated to be $6.5\% \times 
A_{LL}$. The residual transverse polarization component in the interaction 
region is very small (the longitudinal polarization component is 
$>99.8$\%) and the associated effect on the overall scale is smaller than 
$10^{-3} \times A_{LL}$ and is thus negligible for this analysis.

%%%%%%%%%%%%%%%%%%%%%%%%%%%%%%%%%%%%%%%%%%%%%%%%%%%%%%%%%%%%%%  Table_III
\begin{table*}[!hbt]
\caption{\label{tab:result}
\jpsiall as a function of \pt or \absy. $N_{J/\psi}^{2\sigma}$ is the 
\jpsi counting within its $2\sigma$ mass window. The column of Type A 
systematic uncertainties are a statistically weighted quadratic 
combination of the background fraction and run grouping uncertainties. 
$\Delta A_{LL}$ (Rel. Lumi.) is the global systematic uncertainty from 
relative luminosity measurements. $\Delta A_{LL}$ (Polarization) is the 
systematic uncertainty from the beam polarization measurement.}
\begin{ruledtabular}  \begin{tabular}{ccccccccc}
\specialcell[c]{\pt (GeV/$c$) \\or \absy bin} & \specialcell[c]{\meanpt (GeV/$c$)\\ or \meanabsy} &  \specialcell[c]{$N_{J/\psi}^{2\sigma}$\\$\times 10000$} &  \jpsiall & \specialcell[c]{$\Delta A_{LL}$\\(stat)} & \specialcell[c]{$\Delta A_{LL}$\\(Type A syst)} & \specialcell[c]{$\Delta A_{LL}$ (Rel. Lumi.)\\(Type B syst)} & \specialcell[c]{$\Delta A_{LL}$ (Polarization)\\(Type B syst)}\\\hline
\specialcell[c]{\pt $\in$ (0--10)\\\absy $\in$ (1.2--2.2)}  &\specialcell[c]{\meanpt  $=$ 2.03\,GeV/$c$\\ \meanabsy $=$ 1.71}	&15.9 &0.012  &0.010  &0.003  &0.0004  &0.001  \\ \\
\pt $\in$ (0--2)   &1.12  &8.8	&0.003  &0.014  &0.003  &0.0004  &$<$\,0.001  \\
\pt $\in$ (2--4)   &2.79  &5.6	&0.007  &0.016  &0.004  &0.0004  &$<$\,0.001  \\
\pt $\in$ (4--10)  &5.25  &1.7	&0.057  &0.029  &0.010  &0.0004  &0.004  \\
\absy $\in$ (1.2--1.8)   &1.59  &10.2	 &0.025  &0.013  &0.006  &0.0004  &0.002  \\
\absy $\in$ (1.8--2.2)   &1.94  &4.9	 &0.001  &0.019  &0.003  &0.0004  &$<$\,0.001  \\
\end{tabular}  \end{ruledtabular}
\end{table*}

A technique called ``bunch shuffling"~\cite{Adare:2014hsq} was applied to 
test for additional RHIC bunch-to-bunch and fill-to-fill uncorrelated 
systematic uncertainties that may have been overlooked. The resulting 
$A_{LL}^{\rm shuffle}$ follows a Gaussian distribution with $\sigma$ 
consistent with the statistical uncertainty of \jpsiall obtained with real 
data. This test result indicates that all other uncorrelated 
bunch-to-bunch and fill-to-fill systematic uncertainties are much smaller 
than the statistical uncertainties.

%%%%%%%%%%%%%%%%%%%%%%%%%%%%%%%%%%%%%%
\section{Results and Summary}
%%%%%%%%%%%%%%%%%%%%%%%%%%%%%%%%%%%%%%

\begin{figure}[!htb]
\includegraphics[width=1.0\linewidth]{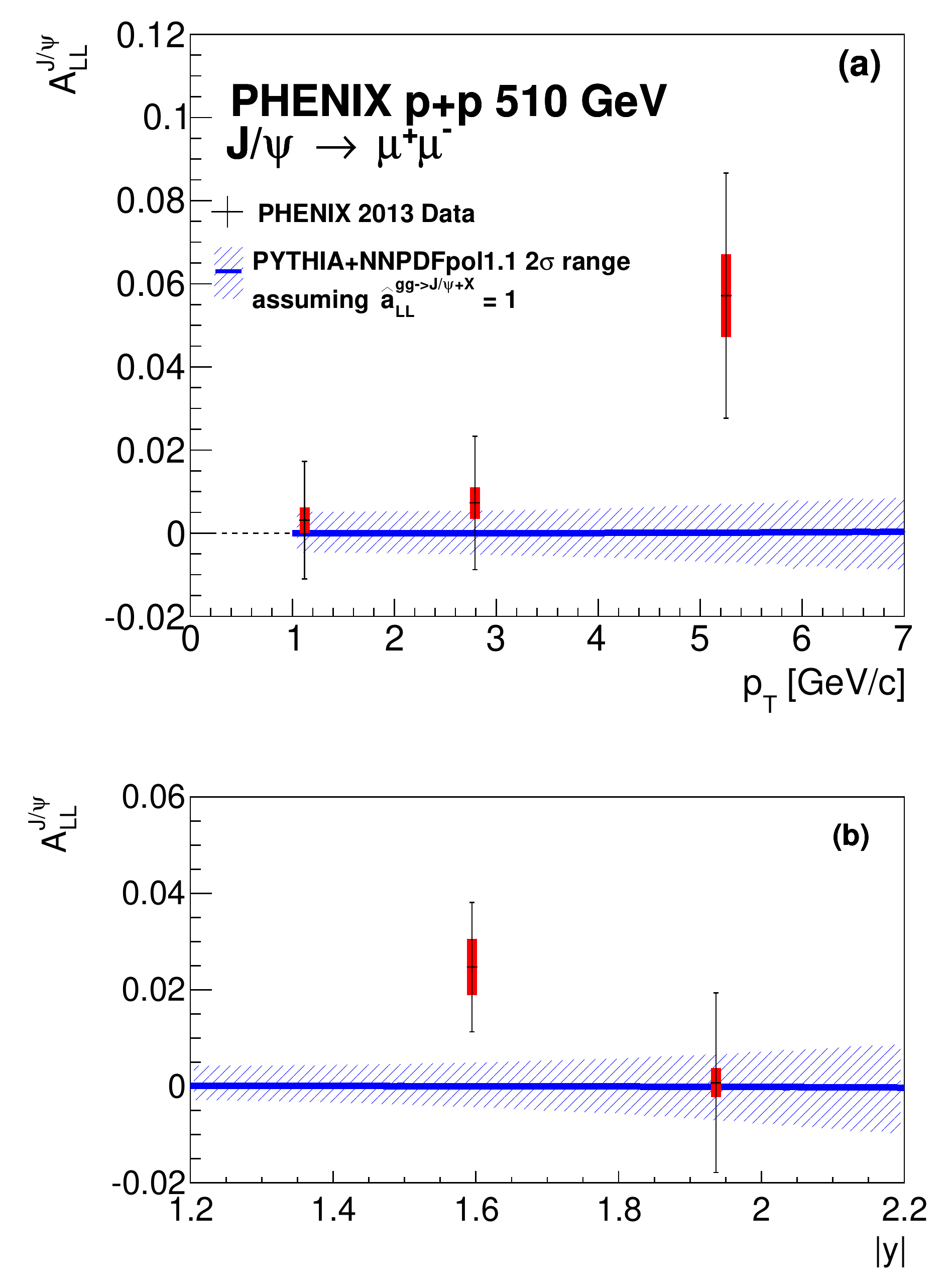}
\caption{\label{fig:A_LL_pT_eta} 
\jpsiall as a function of \pt (top panel) and \absy (bottom panel). The 
black error bars show the statistical uncertainty. The red boxes show only 
the Type A systematic uncertainties. There are additionally a 4\,$\times 
10^{-4}$ global systematic uncertainty from the relative luminosity 
determination and a 6.5\,\% global scaling systematic uncertainty from the 
polarization magnitude determination for all \pt or \absy bins. The blue 
curve with shaded band is our \jpsiall estimation using 
{\sc pythia}6~\cite{Sjostrand:PYTHIA} simulation with NNPDF data sets 
under the assumption of \ahat~$ = 1$. The solid blue curve is the 
central value and the blue shaded band is the $\pm~2~\sigma$ 
uncertainty range. See details in the text. } \end{figure}

The final results for \jpsi \all as a function of \pt and \absy are 
summarized in Table~\ref{tab:result} and in Fig.~\ref{fig:A_LL_pT_eta}. 
The average \jpsiall measured is $0.012 \pm 0.010$\,(stat)\,$\pm 
0.003$(syst).

%%%%%%%%%%%%%%%%%%%%%%%%%%%%%%%%%%%%%%%%%%%%%%%%%%%%%%%%%%%%%%  Fig_3

There were several NRQCD calculations of the \jpsiall for RHIC energies 
$\sqrt{s} =200$ GeV and $\sqrt{s} =500$ GeV ~\cite{Teryaev:1996sr} but 
with the Gehrmann-Stirling and other polarized parton distribution 
functions~\cite{Gehrmann:1995ag} produced in the 1990s. Our knowledge of 
quark and gluon polarizations has been significantly improved over the 
last 10 years ~\cite{deFlorian:2014yva,Nocera:2014gqa}. To 
compare our results with the current understanding of the gluon 
polarization, we have calculated the \jpsiall in our kinematic range using 
a {\sc Pythia}~\cite{Sjostrand:PYTHIA} simulation with 
NNPDFpol1.1~\cite{Nocera:2014gqa} and NNPDF3.0~\cite{Ball:2014uwa} as the 
polarized and unpolarized PDF respectively. To separate the uncertainty 
from the \jpsi production mechanism, we have assumed \ahat = 1, which is 
the leading order partonic asymmetry for open heavy quarks in the heavy 
mass limit at RHIC energies~\cite{Gupta:1997nj}. A $2 \sigma$ uncertainty 
band was also calculated using the replica method as presented in 
Ref.~\cite{Gao:2013bia}. The calculated asymmetry using these assumptions 
is shown in Fig.~\ref{fig:A_LL_pT_eta} together with the PHENIX data. The 
calculated asymmetry is consistent with our data within the statistical 
uncertainties.

%%%%%%%%%%%%%%%%%%%%%%%%%%%%%%%%%%%%%%%%%%%%%%%%%%%%%%%%%%%%%%  Fig_4
\begin{figure}[tbp]
\includegraphics[width=1.0\linewidth]{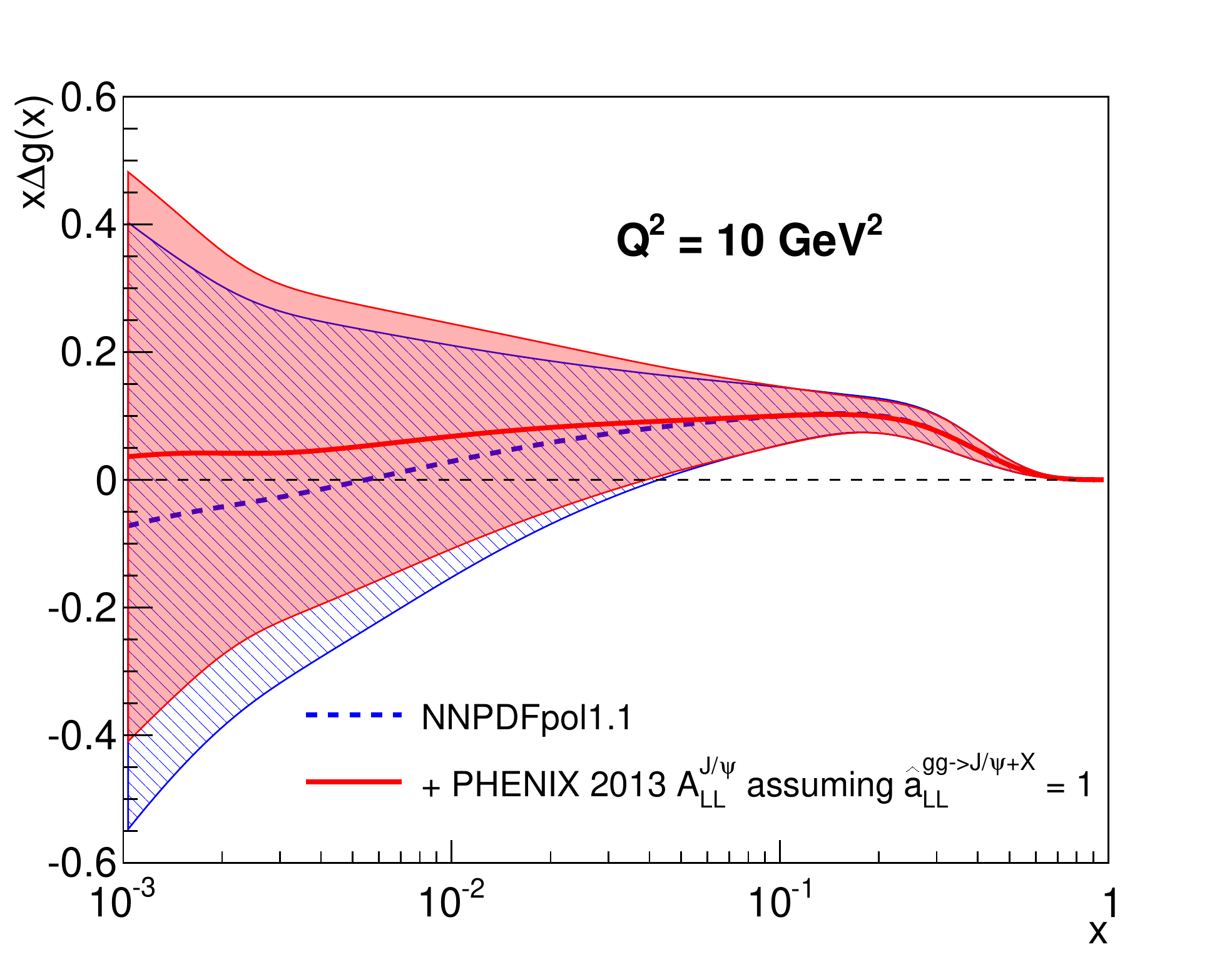}
\caption{\label{fig:impact_plot} 
Blue dashed line is gluon polarization in NNPDFpol1.1; uncertainty band 
for it was obtained from 100 replicas of NNPDFpol1.1 using the replica 
method in~\cite{Gao:2013bia}. Red solid curve is the gluon polarization 
from NNPDFpol1.1 re-weighted using 2013 PHENIX \jpsi \all data under the 
assumption that \ahat = 1.
}
\end{figure}

A reweighting method that estimates the impact of a new dataset on the 
PDFs without doing a new global fit was introduced by the NNPDF
Collaboration~\cite{Ball:2010gb}. Using this method we estimated the 
impact of our data on the gluon polarization based on NNPDFpol1.1 and 
under the assumption of \ahat = 1. Fig.~\ref{fig:impact_plot} shows the 
gluon polarization before and after re-weighting. In this re-weighting, 
only the statistical uncertainty of our data was considered. Under this 
assumption, our data favors a more positive gluon polarization in the $x 
\sim 2 \times 10^{-3}$ region compared to the original NNPDFpol1.1.

In summary, the double helicity asymmetries of inclusive \jpsi production 
have been measured with the PHENIX detector as a function of the \jpsi's 
\pt and \absy, covering $0<\pt<10$~GeV and rapidity \absyrange. The 
\jpsiall measurements offer a new way to access $\Delta G$ via heavy-quark 
production in \pp collisions. They also serve as an important test of the 
universality of the helicity-dependent parton densities and QCD 
factorizations.

%%%%%%%%%%%%%%%%%%%%%%%%%  Acknowledgements

We thank the staff of the Collider-Accelerator and Physics 
Departments at Brookhaven National Laboratory and the staff of 
the other PHENIX participating institutions for their vital 
contributions.  We also thank E.~R.~Nocera for helpful
discussions on the reweighting using NNPDFpol.  We 
acknowledge support from the Office of Nuclear Physics in the
Office of Science of the Department of Energy,
the National Science Foundation,
Abilene Christian University Research Council,
Research Foundation of SUNY, and
Dean of the College of Arts and Sciences, Vanderbilt University
(U.S.A),
Ministry of Education, Culture, Sports, Science, and Technology
and the Japan Society for the Promotion of Science (Japan),
Conselho Nacional de Desenvolvimento Cient\'{\i}fico e
Tecnol{\'o}gico and Funda\c c{\~a}o de Amparo {\`a} Pesquisa do
Estado de S{\~a}o Paulo (Brazil),
Natural Science Foundation of China (P.~R.~China),
Croatian Science Foundation and
Ministry of Science, Education, and Sports (Croatia),
Ministry of Education, Youth and Sports (Czech Republic),
Centre National de la Recherche Scientifique, Commissariat
{\`a} l'{\'E}nergie Atomique, and Institut National de Physique
Nucl{\'e}aire et de Physique des Particules (France),
Bundesministerium f\"ur Bildung und Forschung, Deutscher
Akademischer Austausch Dienst, and Alexander von Humboldt Stiftung (Germany),
National Science Fund, OTKA, K\'aroly R\'obert University College,
and the Ch. Simonyi Fund (Hungary),
Department of Atomic Energy and Department of Science and Technology (India),
Israel Science Foundation (Israel),
Basic Science Research Program through NRF of the Ministry of Education (Korea),
Physics Department, Lahore University of Management Sciences (Pakistan),
Ministry of Education and Science, Russian Academy of Sciences,
Federal Agency of Atomic Energy (Russia),
VR and Wallenberg Foundation (Sweden),
the U.S. Civilian Research and Development Foundation for the
Independent States of the Former Soviet Union,
the Hungarian American Enterprise Scholarship Fund,
and the US-Israel Binational Science Foundation.

%%%%%%%%%%%%%%%%%%%%%%%%%%%  References

%\bibliography{ppg180x1}

%merlin.mbs apsrev4-1.bst 2010-07-25 4.21a (PWD, AO, DPC) hacked
%Control: key (0)
%Control: author (0) dotless jnrlst
%Control: editor formatted (1) identically to author
%Control: production of article title (0) allowed
%Control: page (1) range
%Control: year (0) verbatim
%Control: production of eprint (0) enabled
%
 
\end{document}